\documentclass[conference]{IEEEtran}
\IEEEoverridecommandlockouts
\usepackage{cite}
\usepackage{amsmath,amssymb,amsfonts}

\usepackage{graphicx}
\usepackage{textcomp}
\usepackage{xcolor}
\usepackage{caption}
\usepackage{subcaption}
\usepackage[ruled,vlined]{algorithm2e}
\usepackage{multirow} 
\usepackage{tikz}
\usepackage{pgfplots}
\pgfplotsset{compat=1.18}
\usepackage{algpseudocode}
\usepackage{hyperref}
\def\BibTeX{{\rm B\kern-.05em{\sc i\kern-.025em b}\kern-.08em
    T\kern-.1667em\lower.7ex\hbox{E}\kern-.125emX}}
\begin{document}
\newcommand{\todo}[1]{{\color{red} #1}}
\newcommand{\ytt}[1]{{\color{blue} #1}}
\title{\textit{AXON}: An Automated Netlist Optimization Framework for High-Speed Adders}


\author{
Tiantian Yang,
Xuanle Ren\textsuperscript{*},
Qingdian Wan,
Qi Meng \\
ShanghaiTech University, Shanghai, China \\
{\{yangtt2023, renxl, wanqd2025, mengqi2025\}@shanghaitech.edu.cn}
\thanks{\textsuperscript{*} Corresponding author.}
}

\maketitle

\begin{abstract}
Adders are fundamental building blocks in modern digital systems, such as CPUs and accelerators, and their performance, power, and area (PPA) directly affect overall system efficiency. 
{Contemporary adders commonly adopt parallel-prefix architectures with well-known PPA trade-offs.
However, for specific design objectives, these classical architectures often fail to achieve globally optimal PPA. }
Prior work often optimizes topology without netlist-/cell-level awareness, and general-purpose synthesis heuristics are not adder-specific, leading to suboptimal PPA.
To address these limitations, we propose an automated netlist optimization framework for adders, named AXON. The framework performs comprehensive design space exploration from the architectural to the netlist level, integrating prefix topology search with standard-cell-aware mapping strategies. It employs a hierarchical search methodology that rapidly converges to near-optimal solutions under given PPA constraints. Furthermore, we introduce a hybrid ultra-high-speed adder structure that combines the advantages of the parallel-prefix and Ling architectures to effectively shorten the critical path.
Experimental results based on the TSMC 28nm library show that AXON achieves up to 10.3\%, 12.6\%, and 32.1\% improvement in delay, area–delay product, and energy–delay product, respectively, compared to the synthesis results using commercial synthesis tools.
\end{abstract}

\section{Introduction}\label{sec:introduction}
Adders are among the most fundamental and frequently used arithmetic units in digital systems, directly affecting the overall speed, energy efficiency, and area of a system. In CPUs, digital signal processors (DSPs), and accelerators, adders often reside on the critical path and therefore play a decisive role in determining clock frequency. Different applications impose distinct design requirements: CPUs and GPUs typically demand extremely high operating frequencies to maximize performance, whereas embedded systems prioritize overall power–performance–area (PPA) efficiency. Consequently, meeting specific PPA constraints remains a central objective in adder design.

Most modern adders employ the \textit{parallel-prefix} architecture, which significantly reduces logic depth—and thus computation delay—by parallelizing carry generation and propagation. The \textit{prefix topology}, i.e., the interconnection structure for carry computation, directly determines performance and implementation efficiency. Since the introduction of classical architectures such as Kogge–Stone~\cite{Kogge1973}, Brent–Kung~\cite{Brent1982}, and Sklansky~\cite{Sklansky1960}, numerous studies have explored ways to refine prefix topologies to achieve better trade-offs among delay, area, and routing complexity. Although widely adopted in both academic and industrial designs, these architectures are typically heuristic or experience-driven, and thus cannot guarantee optimal results under specific PPA constraints. For a given technology node, target frequency, and power budget, such fixed prefix structures may be suboptimal.

In practice, adder optimization often requires iterative manual tuning and substantial design expertise. Design teams typically customize parallel-prefix structures for different bit-widths and performance targets. For example, low-power designs may adopt deep and sparse prefix networks to reduce switching activity, while high-performance designs employ shallow and dense networks to minimize delay. This manual process is labor-intensive, time-consuming, and heavily reliant on designer experience, making it difficult to ensure globally optimal PPA outcomes. Furthermore, the number of possible prefix topologies grows exponentially with bit-width~\cite{Roy2016}, rendering exhaustive search or purely heuristic selection infeasible and limiting both design efficiency and solution quality.

To address these challenges, several academic studies have investigated automated optimization techniques. Some use graph-theoretic or heuristic algorithms to optimize prefix topologies with respect to node count and depth~\cite{Roy2014, Roy2016}, while others leverage machine learning to accelerate or guide architectural exploration~\cite{Ma2019, Geng2022, Roy2021}. However, these approaches typically focus solely on prefix topology optimization and neglect \textit{post-mapped netlist characteristics}. In practice, adder performance depends not only on the algorithmic computation of carry signals, but also on synthesis mapping strategies and physical placement/routing effects. Consequently, optimizing the prefix topology alone—without considering technology mapping—cannot yield globally optimal PPA results.

Meanwhile, industrial electronic design automation (EDA) flows rely heavily on commercial synthesis tools such as \textit{Synopsys Design Compiler}~\cite{SynopsysDC} and \textit{Cadence Genus}~\cite{CadenceGenus}. These tools employ general-purpose optimization algorithms for arbitrary logic circuits, rather than domain-specific strategies tailored for adders. As a result, adders synthesized through such tools may not fully exploit the structural advantages of parallel-prefix architectures and often fail to achieve globally optimal PPA.

To overcome these limitations, we propose an automated netlist optimization framework for high-speed adders, named \textit{AXON}, which enables collaborative optimization from the architectural level down to the netlist level. The framework automatically generates adder netlists that satisfy given PPA constraints. It first performs a global architectural search to identify optimal prefix topologies. It then converts the obtained prefix topology into a standard-cell-level netlist using techniques such as hybrid prefix architecture, propagation-signal and inverter insertion, and logic gate resizing. During this process, a structured design space is constructed, supporting efficient standard-cell-aware mapping and performance optimization. Finally, \textit{AXON} explores this design space to identify the optimal netlist using a coarse-to-fine search strategy.
The main contributions of this work are summarized as follows:

\begin{enumerate}
\item We propose an automated, standard-cell-aware netlist optimization framework for high-speed adders that achieves full-flow optimization from architecture to netlist. Notably, although the evaluation is based on a specific technology, the framework is technology-agnostic and portable across different standard-cell libraries.
\item We introduce a hybrid high-speed parallel-prefix architecture that combines the prefix generation mechanisms~\cite{BeaumontSmith2001} and Ling~\cite{Ling1981}. The hybrid structure effectively reduces logic depth and thus computation delay at the cost of a slight area overhead, making it particularly suitable for high-speed designs.
\item We construct a design space for potential adder netlists and employ a coarse-to-fine search strategy to efficiently explore a large number of candidate netlists, significantly improving the efficiency of design space exploration.
\end{enumerate}
\begin{figure}[!t]
\centering

\begin{subfigure}[t]{0.48\columnwidth}
  \centering
  \includegraphics[width=\linewidth]{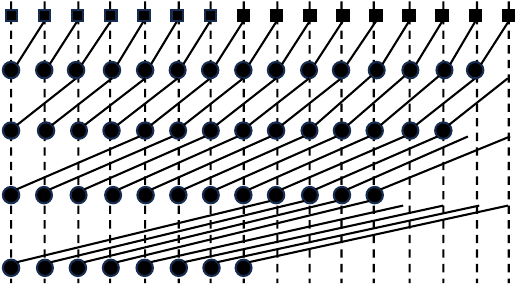}
  \caption{Kogge-Stone}
  \label{fig:ks}
\end{subfigure}\hfill
\begin{subfigure}[t]{0.48\columnwidth}
  \centering
  \includegraphics[width=\linewidth]{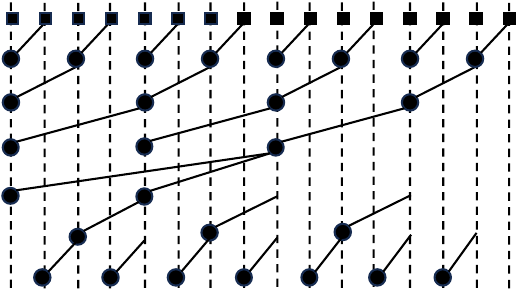}
  \caption{Brent-Kung}
  \label{fig:bk}
\end{subfigure}

\par\vspace{1.5pt}
\begin{subfigure}[t]{0.47\columnwidth}
  \centering
  \includegraphics[width=\linewidth]{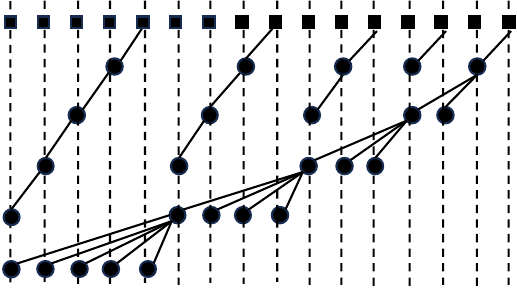}
  \caption{Roy \textit{et al.}~\cite{Roy2014,Roy2016}}
  \label{fig:roy}
\end{subfigure}\hfill
\begin{subfigure}[t]{0.48\columnwidth}
  \centering
  \includegraphics[width=\linewidth]{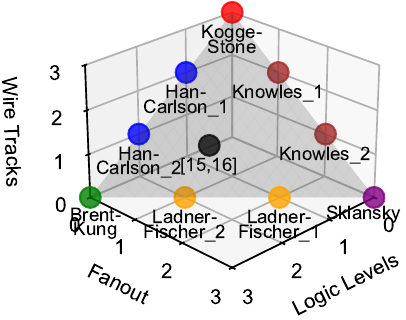}
  \caption{Comparison}
  \label{fig:comp}
\end{subfigure}

\caption{(a)–(c) Prefix topologies of representative adders. (d) Comparison of different adders (adapted from~\cite{Weste2004}).}
\label{fig:ppa-tree}
\vspace{-15pt} 
\end{figure}

\section{Parallel Prefix Adders and Existing Optimization Techniques}\label{sec:background}

This section reviews the fundamentals of parallel-prefix adders and summarizes representative optimization techniques.

\subsection{Parallel Prefix Adders}

\setlength{\abovedisplayskip}{1pt}
\setlength{\belowdisplayskip}{1pt}

Parallel-prefix adders achieve logarithmic-depth carry computation by parallelizing carry generation and propagation~\cite{BeaumontSmith2001,Kogge1973,Brent1982,Sklansky1960,Ling1981,HanCarlson1987}, transforming sequential carry propagation into tree-based prefix computation and reducing the critical path delay from $O(n)$ to $O(\log n)$.

For an $n$-bit addition $A + B$, the carry bit $c_i$ is computed as:
\begin{equation}\label{eq:wein_carry}
c_i = g_i + p_i \cdot c_{i-1}, \quad i = 0,1,\dots,n-1,
\end{equation}
where
\begin{equation}\label{eq:wein_gp}
g_i = A_i \cdot B_i, \quad p_i = A_i \oplus B_i,
\end{equation}
represent the \textit{generate} and \textit{propagate} signals.
{In the prefix network, \((G,P)\) pairs are recursively combined over intervals, where \(G_{i:j}\) and \(P_{i:j}\) denote the group-generate and group-propagate signals on \([i,j]\). The prefix operation is defined as}
\begin{equation}\label{eq:wein_recursion}
G_{i:j} = G_{i:k+1} + P_{i:k+1} \cdot G_{k:j},
\end{equation}
\begin{equation}
P_{i:j} = P_{i:k+1} \cdot P_{k:j},
\end{equation}
yielding
\begin{equation}
c_i = G_{i-1:0} + P_{i-1:0} \cdot c_0,
\end{equation}
and
\begin{equation}
S_i = p_i \oplus c_i.
\end{equation}
Classical parallel-prefix architectures trade off delay, area, and wiring complexity (Fig.~\ref{fig:comp}).
\textit{Kogge--Stone} adders (Fig.~\ref{fig:ks}) use a full binary tree with minimal depth and regular layout, achieving low delay but higher area and routing complexity~\cite{Kogge1973}.
\textit{Brent--Kung} adders (Fig.~\ref{fig:bk}) adopt a folded topology with slightly larger depth but reduced area and power~\cite{Brent1982}.
\textit{Sklansky} adders minimize logic depth via recursive grouping at the cost of routing congestion~\cite{Sklansky1960}.
\textit{Han--Carlson} adders~\cite{HanCarlson1987} balance the above structures for medium bit-widths, and \textit{Ling} adders reduce critical-path delay for specific bit-widths~\cite{Ling1981}.
{These architectures are heuristic and cover only a few discrete points in the vast design space (Fig.~\ref{fig:comp}); thus, for specific PPA constraints, more efficient designs may exist beyond these classical choices.}

\subsection{Existing PPA Design and Optimization Techniques}
Beyond classical topologies (e.g., Kogge--Stone and Brent--Kung), the prefix topology forms a large combinational space, motivating various search and optimization approaches.
Roy \textit{et al.}~\cite{Roy2014} proposed a bottom-up synthesis method and proved size optimality under minimum logic depth; later they derived a polynomial-time algorithm with $O(n^2 \log_2 n)$ complexity under fan-out constraints and improved efficiency via topology-aware node cloning~\cite{Roy2016}, yielding fewer-node topologies (Fig.~\ref{fig:roy}).
Moto \textit{et al.}~\cite{Moto2018} explored the space using simulated annealing with ``prefix sequence'' encoding.
These methods can produce near-optimal topologies but typically rely on commercial EDA tools for netlist synthesis, leaving mapping unoptimized.


{Given the high computational cost of exhaustive EDA evaluations, learning-based methods have been used for faster exploration.
Ma \textit{et al.}~\cite{Ma2019} applied GP-based active learning to discover Pareto-optimal designs from structural features.
Geng \textit{et al.}~\cite{Geng2022} proposed a Graph Neural Process model (VGAE + neural process) for end-to-end learning and sequence optimization, improving Pareto fronts.
PrefixRL~\cite{Roy2021} uses reinforcement learning to explore unconstrained prefix spaces and reports up to 30.2\% area reduction for 64-bit adders, though learning-based approaches may require additional verification.

Hybrid adders have also been explored, including the Ling adder by Keeter \textit{et al.}~\cite{Keeter2011} and a CSA--CLA--RCA hybrid by Deshmukh and Mhala~\cite{DeshmukhHybridAdder}, both improving performance and power efficiency.}

\section{Methodology}\label{sec:methodology}

This section presents the design methodology and optimization workflow of the \textit{AXON} framework.

\begin{figure}[t]
    \centering
    \includegraphics[width=0.5\textwidth]{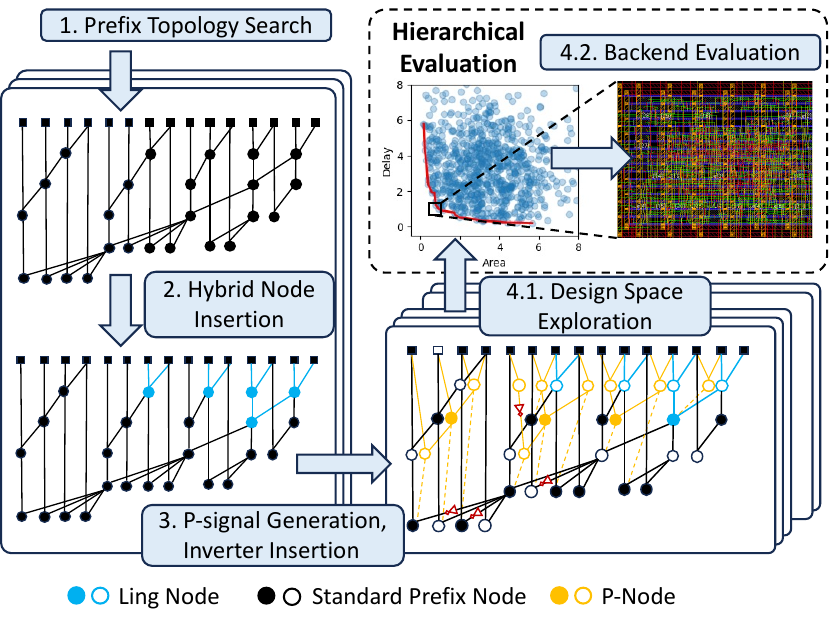}
    \caption{\textit{AXON} Overview.}
    \label{fig:axon}
    \vspace{-5pt} 
\end{figure}

\subsection{\textit{AXON} Overview}\label{sec:axon_framework}

As illustrated in Fig.~\ref{fig:axon}, \textit{AXON} is a hierarchical, automated netlist optimization framework targeting high-speed adders. The optimization workflow of \textit{AXON} consists of four stages:

\begin{enumerate}
{\item \textbf{Prefix Topology Search:} We start from a depth-first search procedure based on Roy et al..~\cite{Roy2016} to obtain a minimal-node prefix topology under maximum fanout constraints. The logical depth is configured according to the target delay.}

\item \textbf{Hybrid Node Insertion:} Ling nodes are selectively inserted along the critical path and combined with prefix nodes to reduce logic levels in the prefix topology, thereby decreasing critical-path delay.

\item \textbf{Propagate Signal Generation and Inverter Insertion:} The propagate ($P$) network is constructed on demand, and inverters are inserted to align signal polarity and optimize gate driving strength. Multi-fanout units are analyzed using the \textit{logical effort} method, and gate sizing is adjusted to optimize global delay, generating an expanded design space for subsequent exploration.

\item \textbf{Hierarchical Evaluation:} A two-stage evaluation is applied: (i) fast logic-level estimation for coarse-grained delay and area, and (ii) full physical-level evaluation using commercial tools for accurate post-layout timing and area metrics.
\end{enumerate}

\begin{figure}[t]
    \centering
    \includegraphics[width=0.42\textwidth]{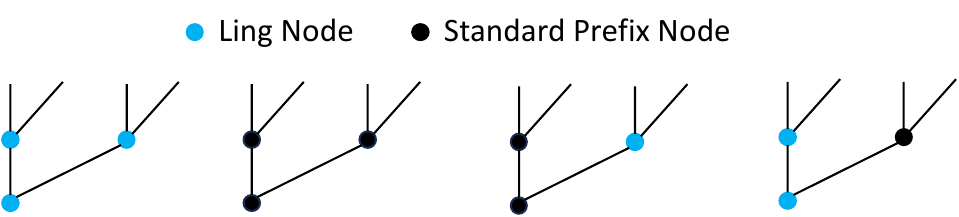}
    \caption{Conversion between standard prefix nodes and Ling nodes.}
    \label{fig:wein_ling}
    \vspace{-8pt}
\end{figure}

\subsection{Hybrid Parallel-Prefix–Ling Adder Architecture}\label{sec:wein_ling}

\setlength{\abovedisplayskip}{1pt}
\setlength{\belowdisplayskip}{1pt}

Unlike the standard parallel-prefix adder, which uses the prefix signals defined in Eq.~\ref{eq:wein_gp}, the Ling adder introduces a different definition for the propagate signal:
\begin{equation}\label{eq:ling_gp}
g_i = A_i \cdot B_i, \quad p_i = A_i + B_i,
\end{equation}
leading to the carry computation:
\begin{equation}\label{eq:ling_rec}
c_i = g_i + p_i g_{i-1}.
\end{equation}
A pseudo-carry signal $H_i$ is further defined in the Ling formulation:
\begin{equation}\label{eq:ling_carry}
H_i = c_i + c_{i-1}.
\end{equation}
Using Eqs.~\ref{eq:ling_rec} and \ref{eq:ling_carry}, the recursion of $H_i$ becomes:
\begin{equation}
H_{i:0} = g_i + g_{i-1} + p_{i-1}g_{i-2} + p_{i-1}p_{i-2} \dots p_{1} g_0,
\end{equation}
and more generally:
\begin{equation}\label{eq:ling_recursion}
H_{i:j} = H_{i:k} + P_{i-1:k} H_{k-1:0}.
\end{equation}

The Ling recursion is analogous to the recursion of the parallel-prefix topology, with the primary difference being the range of the $p$ signals. Ling nodes can also be converted to standard prefix nodes:
\begin{equation}
G_{i:j} = p_{i} \cdot H_{i:j}.
\end{equation}
Thus, within the prefix topology, Ling nodes and standard prefix nodes can be interconverted, as shown in Fig.~\ref{fig:wein_ling} and Eqs.~\ref{eq:ling2wein}–\ref{eq:wein2ling}:
\begin{equation}\label{eq:ling2wein}
G_{i:j} = G_{i:k} + P_{i:k} H_{k-1:0},
\end{equation}
\begin{equation}\label{eq:wein2ling}
H_{i:j} = H_{i:k} + P_{i-1:k} G_{k-1:0}.
\end{equation}

The introduction of the Ling architecture reduces logic depth within the critical path of the parallel-prefix topology. For example, $H_{1:0} = g_1 + g_0 = A_1B_1 + A_0B_0$ can be implemented using a single AOI22 gate, whereas $G_{1:0} = g_1 + p_1 g_0 = A_1B_1 + (A_1 \oplus B_1)A_0B_0$ requires much more logic levels. The trade-off is a slight increase in area due to the need to generate two types of propagate signals (Eqs.~\ref{eq:wein_gp} and~\ref{eq:ling_gp}). 
Therefore, \textit{AXON} {first compute node arrival times under a coarse delay model} and then applies Ling node transformation only along the critical path, while non-critical paths retain the standard prefix nodes to maintain area efficiency.
In summary, Ling nodes effectively reduce delay on the critical path of the prefix topology, whereas standard prefix nodes preserve structural compactness in non-critical regions. The hybrid deployment of these node types in \textit{AXON} achieves an adaptive balance between performance and area.

\subsection{Insertion of Propagate Gates and Inverters}\label{sec:p_network}

{In parallel-prefix adders, carries are computed by recursively combining Generate ($G$) and Propagate ($P$) signals. While structured topologies (e.g., Kogge--Stone, Brent--Kung) admit regular $P$ construction, irregular and hybrid (prefix--Ling) topologies complicate $P$ generation because the valid $P$ spans differ across node types (Eqs.~\ref{eq:wein_recursion},~\ref{eq:ling_recursion}, and~\ref{eq:ling2wein}--\ref{eq:wein2ling}).}

{Generating $P_{i:j}$ alongside all $G_{i:j}$ introduces redundancy, as many $P$ nodes are never used in carry computation. \textit{AXON} therefore builds the $P$ network in a demand-driven manner: starting from required output nodes, it performs bottom-up dependency analysis and instantiates $P_{i:j}$ only when needed by its parents, reducing gate count and routing overhead.}

After the $P$ network is constructed, the combined $P$--$G$ network is mapped to the target standard-cell library. Because CMOS cells often embed inversion functionality (e.g., NAND2 versus AND2), signal polarity naturally alternates across prefix levels, and an inverter is required when connected nodes demand different polarities.

\begin{figure}[t]
    \centering
    \subfloat[\label{fig:inv1}]{
        \includegraphics[width=0.42\columnwidth]{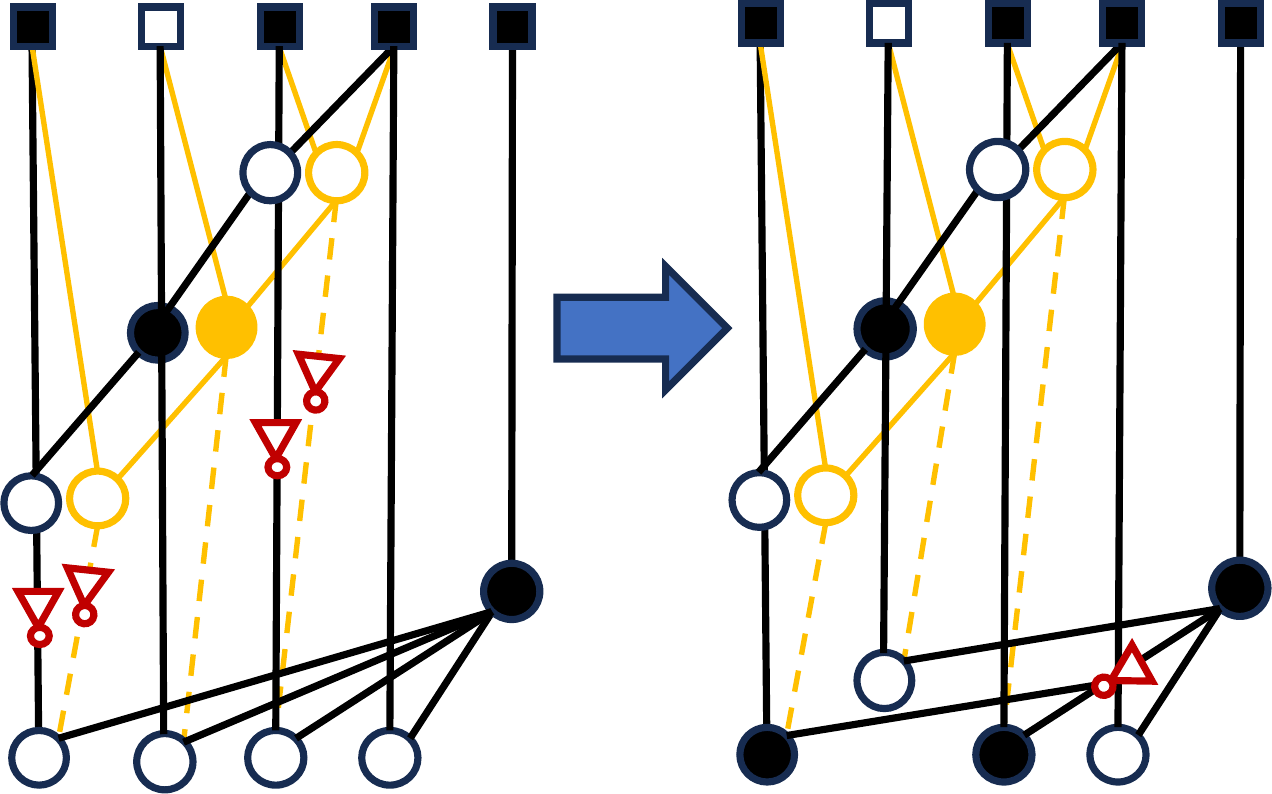}
    }\hfill
    \subfloat[\label{fig:inv2}]{
        \includegraphics[width=0.46\columnwidth]{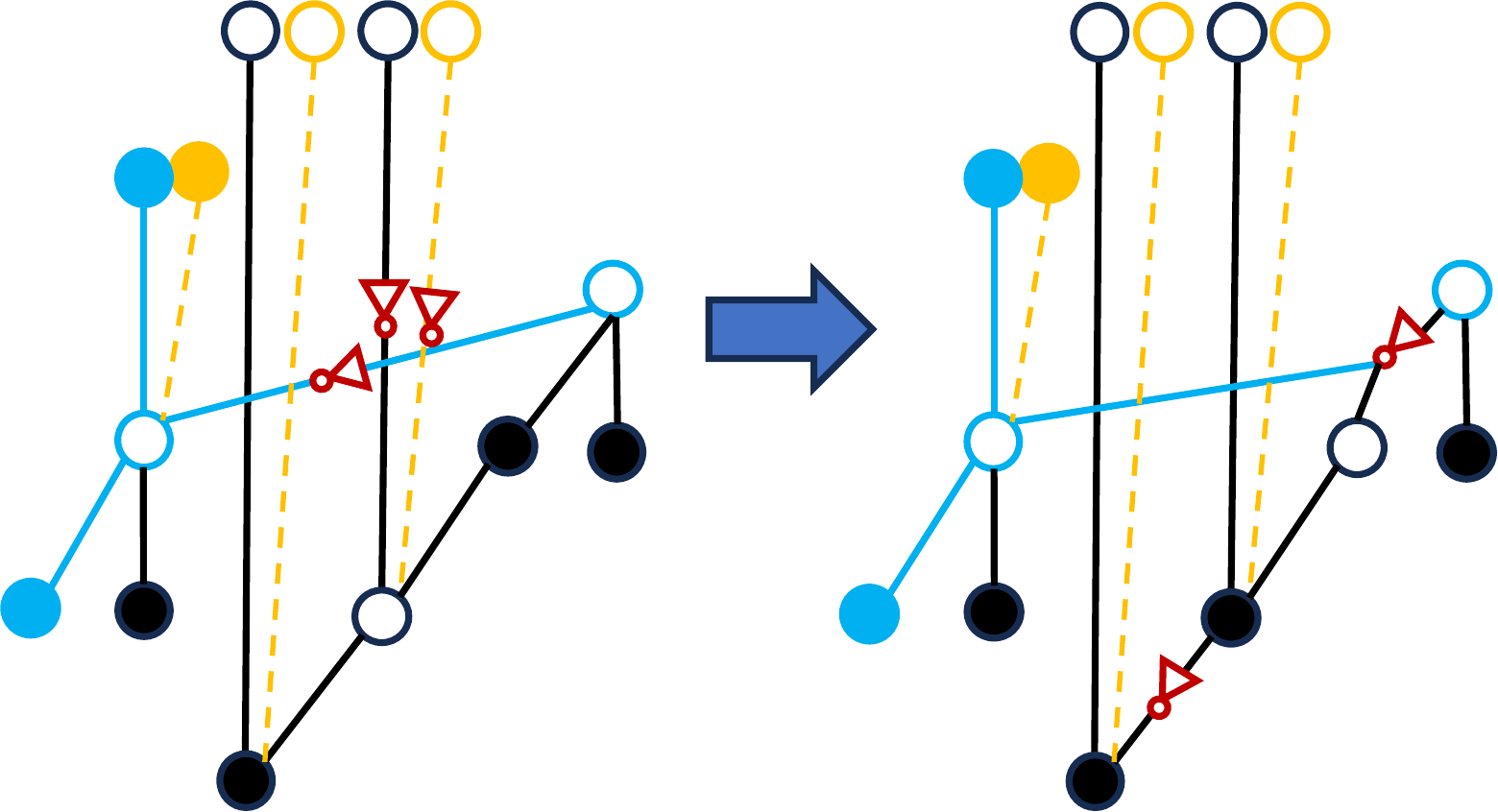}
    }
    \caption{Different inverter insertion strategies. Solid nodes represent positive polarity, while hollow nodes represent negative polarity. If nodes of the same polarity are connected, it means an inverter needs to be inserted on that edge. This figure shows reducing the number of inverters may increase the critical path.}
    \label{fig:inv}
    \vspace{-5pt}
\end{figure}

\begin{figure}[t]
    \centering
    \includegraphics[width=0.45\textwidth]{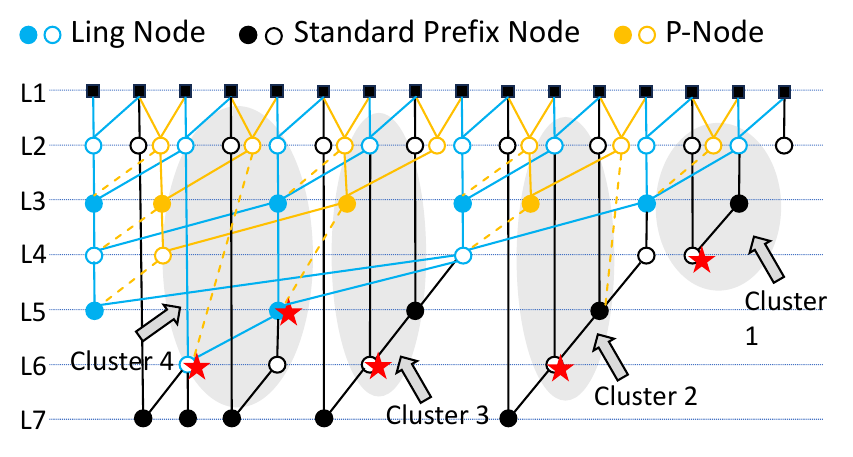}
    \caption{Traversal method for inverter insertion: nodes requiring inversion (marked by stars) are grouped into independent clusters, and inverter positions are traversed within each cluster.}
    \label{fig:traverse}
    \vspace{-6pt} 
    \end{figure}

{In irregular topologies with non-uniform $P$ dependencies, multiple such polarity mismatches may arise. Since different insertion locations lead to different delay--area trade-offs, inverter placement must be optimized. As shown in Fig.~\ref{fig:inv}, fewer inverters do not always improve performance. \textit{AXON} enumerates inverter placements using the traversal procedure in Fig.~\ref{fig:traverse} and Alg.~\ref{alg:inv_insertion}. Two mismatches are grouped into the same cluster if their feasible insertion regions overlap or if resolving one affects the other; otherwise, they are handled independently. This decomposition reduces the global search to several smaller local searches while preserving the key coupling among mismatches.}
\begin{algorithm}[t]
    \caption{Traversal-Based Inverter Insertion}
    \label{alg:inv_insertion}
    \SetAlgoLined
    \SetNlSty{textbf}{}{.}
    \SetAlgoNlRelativeSize{-1}
    \KwIn{Prefix topology $G=(V,E)$, standard-cell library $L$}
    \KwOut{Candidate netlists with all feasible inverter placements}
    
    \BlankLine
    Assign node polarity by level: odd $\rightarrow$ positive, even $\rightarrow$ negative;
    
    \BlankLine
    \ForEach{$(u,v)\in E$}{
        \If{polarity($u$) = polarity($v$)}{
            mark edge $(u,v)$ as a mismatch;
        }
    }
    
    \BlankLine
    Separate mismatched edges into independent clusters;\\
    Enumerate all feasible inverter-insertion positions within each cluster;\\
    Generate the corresponding candidate netlist for each insertion strategy;\\
    Output all candidate netlists.
\end{algorithm}

\begin{table}[t]
\centering
\caption{Impact of hybrid parallel-prefix and Ling architectures on delay, area, and area--delay product (ADP) across adders of different bit-widths.}
\label{tab:wein_ling_eval}
\footnotesize
\resizebox{\linewidth}{!}{
\begin{tabular}{c|c c c|c c c}
\hline
\multirow{2}{*}{Bits} & \multicolumn{3}{c|}{Parallel-Prefix-Only} & \multicolumn{3}{c}{Hybrid} \\
\cline{2-7}
& Delay & Area & ADP & Delay & Area & ADP \\
\hline
16 & 0.294 & 59.976  & 17.609 & 0.235 (-20.1\%) & 65.520 (+9.2\%) & 15.397 (-12.6\%) \\     
23 & 0.344 & 94.080  & 32.363 & 0.289 (-16.0\%) & 100.128 (+6.4\%) & 28.937 (-10.6\%) \\     
31 & 0.357 & 134.568 & 48.040 & 0.298 (-16.5\%) & 141.456 (+5.1\%) & 42.154 (-12.3\%) \\
32 & 0.359 & 139.776 & 50.214 & 0.295 (-17.8\%) & 146.496 (+4.8\%) & 43.216 (-13.9\%) \\
\hline
\end{tabular}
}
\footnotesize\textit{Note: Delay in ns, Area in $\mu$m$^2$, ADP = Area $\times$ Delay.}
\vspace{-5pt} 
\end{table}

\subsection{Design Space Exploration}\label{sec:design_space}
\begin{figure*}[t]
\centering

\begin{subfigure}[t]{0.24\textwidth}
  \centering
  \includegraphics[width=\linewidth]{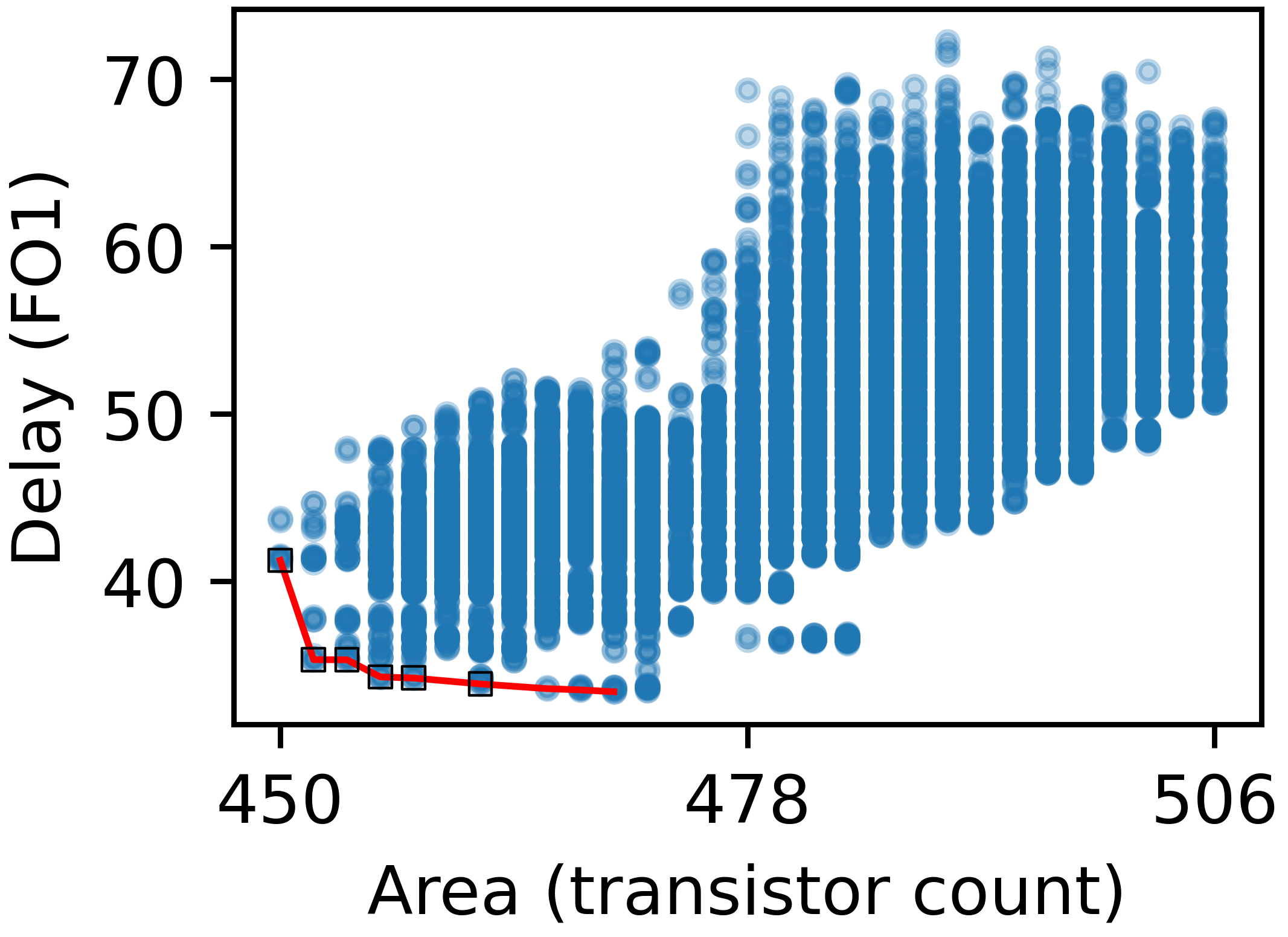}
  \caption{16-bit}
  \label{fig:scatter_16bit}
\end{subfigure}\hfill
\begin{subfigure}[t]{0.24\textwidth}
  \centering
  \includegraphics[width=\linewidth]{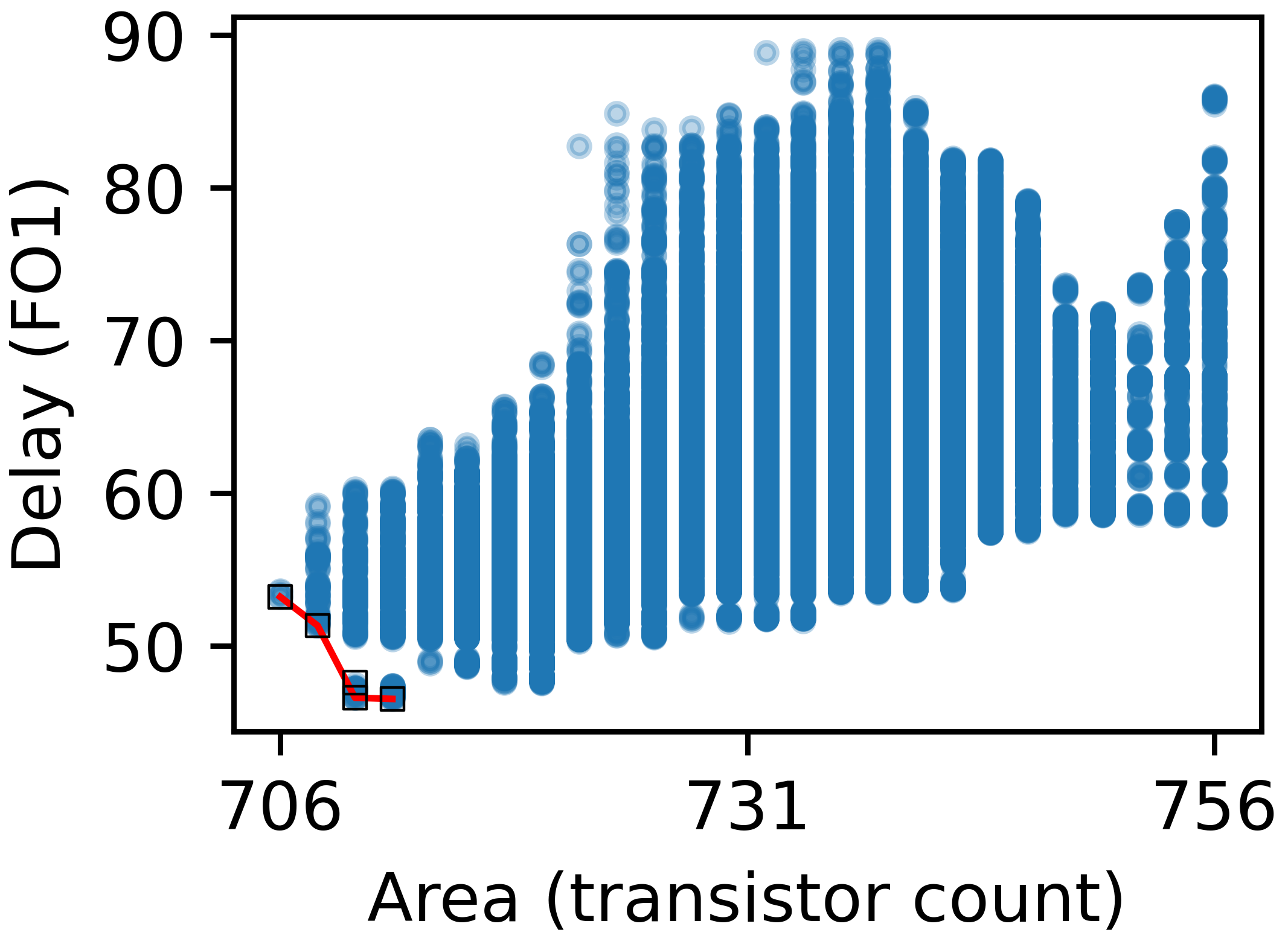}
  \caption{23-bit}
  \label{fig:scatter_23bit}
\end{subfigure}\hfill
\begin{subfigure}[t]{0.24\textwidth}
  \centering
  \includegraphics[width=\linewidth]{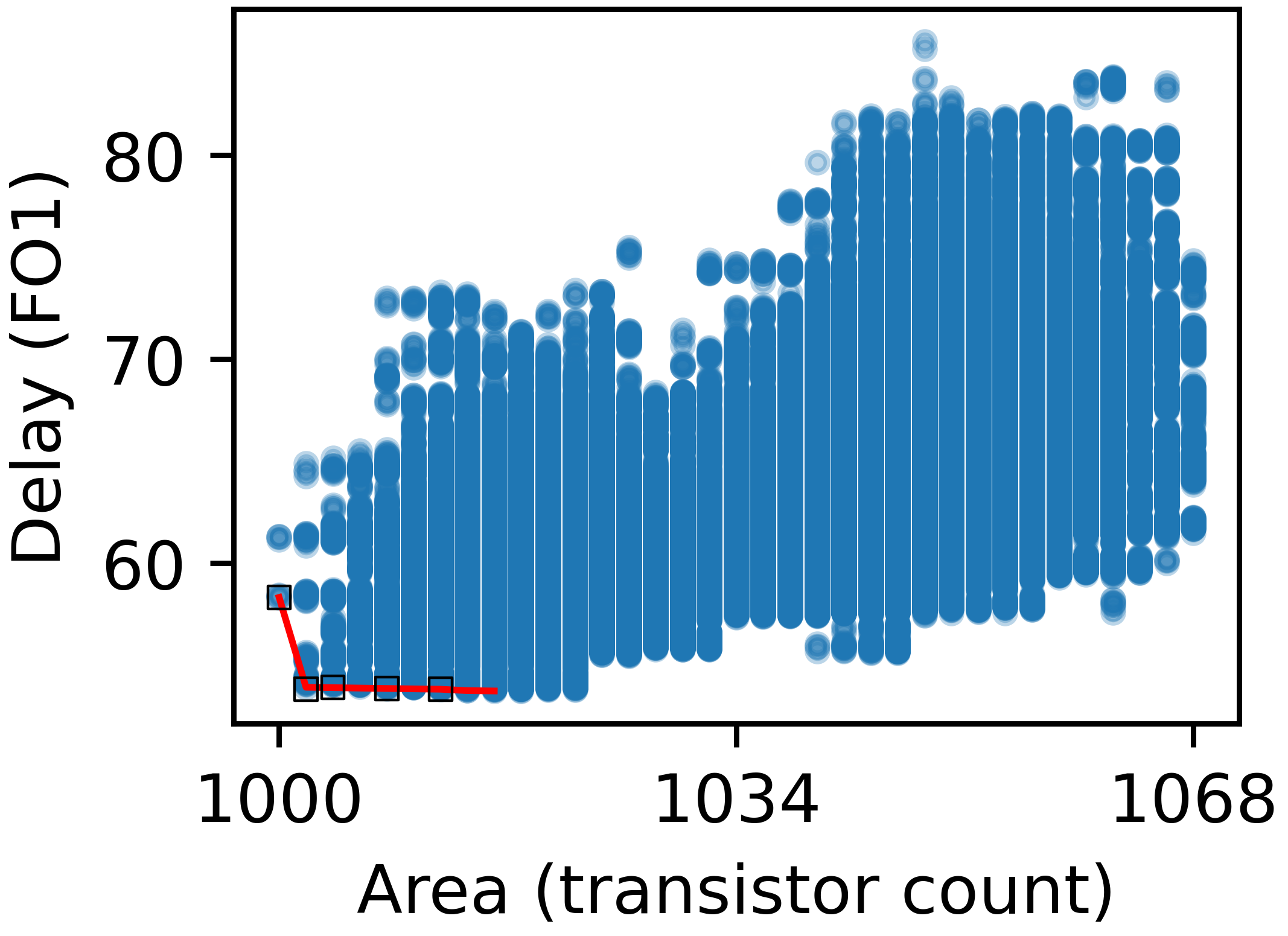}
  \caption{31-bit}
  \label{fig:scatter_31bit}
\end{subfigure}\hfill
\begin{subfigure}[t]{0.24\textwidth}
  \centering
  \includegraphics[width=\linewidth]{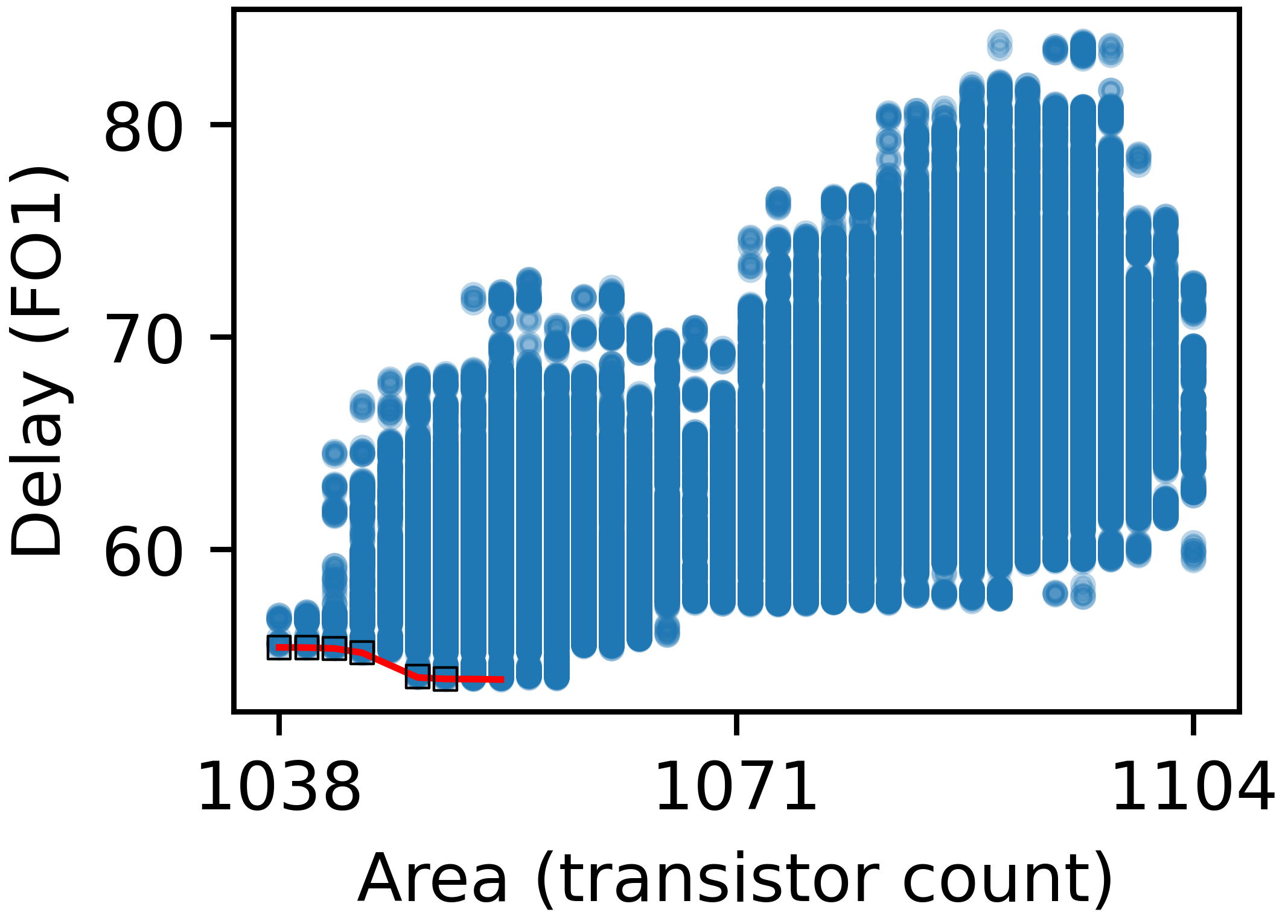}
  \caption{32-bit}
  \label{fig:scatter_32bit}
\end{subfigure}

\caption{Design-space exploration results for four benchmark adders. Each point represents a candidate netlist. Area is measured by transistor count, while delay is measured in units of \textit{FO1}, defined as the delay of a unit-sized inverter driving an identical inverter.}
\label{fig:design_space_scatter}
\vspace{-6pt}
\end{figure*}
\pgfplotsset{compat=1.18}
\begin{figure*}[t]
\centering

\begin{tikzpicture}
\draw[fill=blue!60, draw=black, line width=0.3pt] (0.3,0) rectangle (0.7,0.2);
\node[anchor=west, font=\scriptsize] at (0.7,0.1) {A+B};
\draw[fill=red!60, draw=black, line width=0.3pt] (1.8,0) rectangle (2.2,0.2);
\node[anchor=west, font=\scriptsize] at (2.2,0.1) {Optimal-Prefix-Topology Only~\cite{Roy2014,Roy2016}};
\draw[fill=green!60, draw=black, line width=0.3pt] (6.6,0) rectangle (7.0,0.2);
\node[anchor=west, font=\scriptsize] at (7.0,0.1) {AXON};
\end{tikzpicture}

\vspace{4pt}

\begin{subfigure}[t]{0.25\textwidth}
\centering
\begin{tikzpicture}
\begin{axis}[
    ybar,
    bar width=2.5pt,
    width=\linewidth,
    height=0.75\linewidth,
    ylabel={Delay (ns)},
    ylabel style={font=\scriptsize, yshift=-4pt},
    symbolic x coords={16b+L5,16b+L6,23b+L6,23b+L7,31b+L6,31b+L7,32b+L6,32b+L7},
    xtick=data,
    xticklabel style={font=\scriptsize, rotate=45, anchor=east},
    yticklabel style={font=\scriptsize},
    ymin=0,
    ymax=0.45,
    enlarge x limits=0.05,
]
\addplot[fill=blue!60,bar shift=-\pgfplotbarwidth] coordinates {(16b+L5,0.310) (16b+L6,0.310) (23b+L6,0.372) (23b+L7,0.372) (31b+L6,0.364) (31b+L7,0.364) (32b+L6,0.374) (32b+L7,0.374)};
\addplot[fill=red!60,bar shift=0pt] coordinates {(16b+L5,0.28) (16b+L6,0.28) (23b+L6,0.293) (23b+L7,0.348) (31b+L6,0.324) (31b+L7,0.356) (32b+L6,0.329) (32b+L7,0.354)};
\addplot[fill=green!60, bar shift=\pgfplotbarwidth] coordinates {(16b+L5,0.235) (16b+L6,0.257) (23b+L6,0.289) (23b+L7,0.330) (31b+L6,0.298) (31b+L7,0.328) (32b+L6,0.295) (32b+L7,0.328)};
\end{axis}
\end{tikzpicture}
\caption{Comparison of delay.}
\label{fig:ppa_delay}
\end{subfigure}%
\begin{subfigure}[t]{0.25\textwidth}
\centering
\begin{tikzpicture}
\begin{axis}[
    ybar,
    bar width=2.5pt,
    width=\linewidth,
    height=0.75\linewidth,
    ylabel={Area ($\mu$m$^2$)},
    ylabel style={font=\scriptsize, yshift=-4pt},
    symbolic x coords={16b+L5,16b+L6,23b+L6,23b+L7,31b+L6,31b+L7,32b+L6,32b+L7},
    xtick=data,
    xticklabel style={font=\scriptsize, rotate=45, anchor=east},
    yticklabel style={font=\scriptsize},
    ymin=0,
    ymax=160,
    enlarge x limits=0.05,
]
\addplot[fill=blue!60,bar shift=-\pgfplotbarwidth] coordinates {(16b+L5,66.192) (16b+L6,66.192) (23b+L6,100.632) (23b+L7,100.632) (31b+L6,144.648) (31b+L7,144.648) (32b+L6,149.688) (32b+L7,149.688)};
\addplot[fill=red!60,bar shift=0pt] coordinates {(16b+L5,65.184) (16b+L6,65.184) (23b+L6,103.488) (23b+L7,99.624) (31b+L6,143.808) (31b+L7,142.296) (32b+L6,150.360) (32b+L7,146.664)};
\addplot[fill=green!60,bar shift=\pgfplotbarwidth] coordinates {(16b+L5,65.520) (16b+L6,64.848) (23b+L6,100.128) (23b+L7,93.744) (31b+L6,141.456) (31b+L7,137.592) (32b+L6,146.496) (32b+L7,141.792)};
\end{axis}
\end{tikzpicture}
\caption{Comparison of area.}
\label{fig:ppa_area}
\end{subfigure}%
\begin{subfigure}[t]{0.25\textwidth}
\centering
\begin{tikzpicture}
\begin{axis}[
    ybar,
    bar width=2.5pt,
    width=\linewidth,
    height=0.75\linewidth,
    ylabel={ADP ($\mu$m$^2\!\cdot\!$ns)},
    ylabel style={font=\scriptsize, yshift=-4pt},
    symbolic x coords={16b+L5,16b+L6,23b+L6,23b+L7,31b+L6,31b+L7,32b+L6,32b+L7},
    xtick=data,
    xticklabel style={font=\scriptsize, rotate=45, anchor=east},
    yticklabel style={font=\scriptsize},
    ymin=0,
    ymax=60,
    enlarge x limits=0.05,
]
\addplot[fill=blue!60,bar shift=-\pgfplotbarwidth] coordinates {(16b+L5,20.52) (16b+L6,20.52) (23b+L6,37.44) (23b+L7,37.44) (31b+L6,52.65) (31b+L7,52.65) (32b+L6,55.98) (32b+L7,55.98)};
\addplot[fill=red!60,bar shift=0pt] coordinates {(16b+L5,18.25) (16b+L6,18.25) (23b+L6,30.32) (23b+L7,34.67) (31b+L6,46.59) (31b+L7,50.66) (32b+L6,49.47) (32b+L7,51.92)};
\addplot[fill=green!60,bar shift=\pgfplotbarwidth] coordinates {(16b+L5,15.40) (16b+L6,16.67) (23b+L6,28.94) (23b+L7,30.94) (31b+L6,42.15) (31b+L7,45.13) (32b+L6,43.22) (32b+L7,46.51)};
\end{axis}
\end{tikzpicture}
\caption{Area-delay product.}
\label{fig:ppa_adp}
\end{subfigure}%
\begin{subfigure}[t]{0.25\textwidth}
\centering
\begin{tikzpicture}
\begin{axis}[
    ybar,
    bar width=2.5pt,
    width=\linewidth,
    height=0.75\linewidth,
    ylabel={EDP ($\mu$W$\!\cdot\!$ns$^2$)},
    ylabel style={font=\scriptsize, yshift=-4pt},
    symbolic x coords={16b+L5,16b+L6,23b+L6,23b+L7,31b+L6,31b+L7,32b+L6,32b+L7},
    xtick=data,
    xticklabel style={font=\scriptsize, rotate=45, anchor=east},
    yticklabel style={font=\scriptsize},
    ymin=0,
    ymax=30,
    enlarge x limits=0.05,
]
\addplot[fill=blue!60,bar shift=-\pgfplotbarwidth] coordinates {(16b+L5,7.69) (16b+L6,7.69) (23b+L6,13.29) (23b+L7,13.29) (31b+L6,22.89) (31b+L7,22.89) (32b+L6,25.30) (32b+L7,25.30)};
\addplot[fill=red!60,bar shift=0pt] coordinates {(16b+L5,6.04) (16b+L6,6.04) (23b+L6,10.56) (23b+L7,14.07) (31b+L6,19.00) (31b+L7,22.07) (32b+L6,20.37) (32b+L7,22.57)};
\addplot[fill=green!60,bar shift=\pgfplotbarwidth] coordinates {(16b+L5,3.81) (16b+L6,4.49) (23b+L6,9.10) (23b+L7,11.65) (31b+L6,13.50) (31b+L7,16.78) (32b+L6,13.82) (32b+L7,17.20)};
\end{axis}
\end{tikzpicture}
\caption{Energy-delay product.}
\label{fig:ppa_edp}
\end{subfigure}

\caption{Comparison of area, delay, area-delay-product (ADP), and energy-delay-product (EDP), for \textit{AXON} and two baselines, based on adders with different bit-widths and logic levels.}
\label{fig:ppa_delay_area}
\end{figure*}
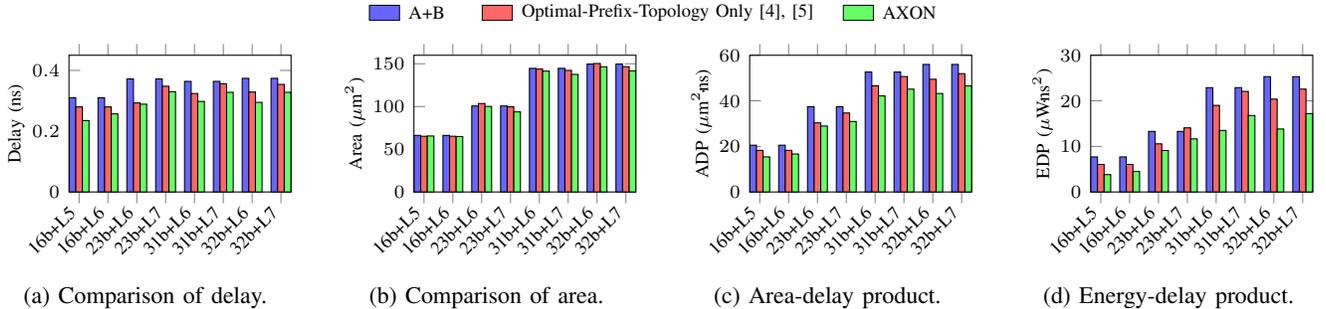

As shown in Fig.~\ref{fig:axon}, both prefix-topology search (Step~1) and $P$-network/inverter inversion (Step~3) produce numerous adder variants. Different prefix topologies and $P$-network/inverter schemes collectively form the \textit{adder design space}. While all candidates implement the same logic, variations in carry propagation and inverter placement lead to differences in gate count, area, and delay.

Full back-end evaluation of all candidates (e.g., placement and routing in commercial tools) is computationally prohibitive. To address this, \textit{AXON} first performs a coarse-grained pre-screening based on delay and area estimates. Area is approximated from transistor counts, and delay is measured using a simplified model adopted from~\cite{Chen2000}. This enables rapid evaluation of tens of thousands of candidates and identifies the promising netlists.

Candidates are visualized in a delay–area scatter plot (as shown in the upper-left scatter plot of Fig.~\ref{fig:axon}), where each point represents a candidate adder. Points closer to the lower-left corner indicate better balance between delay and area. Note that power is not separately evaluated, as it correlates with transistor count and routing. A small subset (typically 5-20) near the Pareto frontier is selected for detailed analysis.

In the second stage, automated place-and-route (APR) provides accurate post-layout delay and area evaluation. This hierarchical two-stage exploration—\textit{generation, fast pre-screening, and precise validation}—efficiently identifies adder netlists with near-optimal delay–area trade-offs.

\section{Evaluation}\label{sec:evaluation}

This section evaluates \textit{AXON} and compares the evaluation results with other approaches.

\subsection{Evaluation of Hybrid Parallel-Prefix and Ling Architecture}

To quantify the impact of introducing Ling nodes along the critical path, we evaluate four adder benchmarks with bit-widths of 16, 23, 31, and 32. All experiments are performed using a \textit{TSMC 28nm} library. The hybrid netlists generated by \textit{AXON} are automatically placed and routed using \textit{Cadence Innovus}~\cite{CadenceInnovus}, and the results are compared against the original parallel-prefix-only counterparts. Evaluation metrics include delay, area, and the area--delay product.

As summarized in Table~\ref{tab:wein_ling_eval}, the incorporation of Ling nodes reduces the critical-path delay by approximately 16\%--20\%, at the cost of a modest area overhead of up to 9.2\% due to the additional transistors required by Ling nodes. This leads to an improvement in the area--delay product (ADP) of approximately 10\%--14\%. The results confirm that augmenting a parallel-prefix adder with Ling-based optimizations not only enhances the speed of the adder but also achieves a more favorable trade-off between area and performance.

\subsection{Design Space Exploration Results}
{We performed the design-space exploration in Sections~\ref{sec:p_network}--\ref{sec:design_space} on four adders (16/23/31/32-bit). For each benchmark, we first searched minimal-node prefix topologies and then generated multiple netlists by traversing $P$-signal construction and inverter insertion, producing thousands to tens of thousands of candidates. Each candidate is evaluated by area (transistor count) and delay using the simplified model from~\cite{Chen2000}:
\begin{equation}
d_{i,j} = d_{\mathrm{int}} + r_{\mathrm{dr}} \cdot C_{\mathrm{load}}
\label{eq:coarse_delay}
\end{equation}
where $d_{\mathrm{int}}$ is intrinsic delay, $r_{\mathrm{dr}}$ is the effective driving length, and $C_{\mathrm{load}}$ is the total input capacitance of fanout gates. Delay is reported in \textit{FO1}, i.e., a unit inverter driving an identical inverter.}


{Fig.~\ref{fig:design_space_scatter} shows that, even with identical bit-width and node count, different $P$-signal and inverter choices lead to large delay/area variations. For the 32-bit benchmark, the maximum area is only 6\% larger than the minimum, while the maximum delay is 56\% higher. We extract a Pareto frontier for each benchmark and select 5--20 candidates near the lower-left region for fine-grained evaluation.}

\subsection{Fine-Grained Evaluation Results}
\begin{figure}[t]
    \centering
    \begin{tikzpicture}
    \begin{axis}[
        width=0.45\textwidth,
        height=0.3\textwidth,
        xlabel={Area ($\mu$m$^2$)},
        ylabel={Delay (ns)},
        grid=major,
        ticklabel style={font=\small},
        title style={font=\small},
        xlabel style={font=\small},
        ylabel style={font=\small},
        nodes near coords style={font=\small}
    ]
    
    \addplot[scatter,black,dashed,
             nodes near coords,
             point meta=explicit symbolic]
    coordinates {
        (157.584, 0.284) []  
        (151.872, 0.308) []  
        (146.664, 0.354) []  
        (143.136, 0.417) []  
        (139.776, 0.479) [Ref~\cite{Roy2014,Roy2016}]  
        (135.408, 0.545) []  
        (131.544, 0.591) []  
    };
    
    \addplot[scatter,only marks,red,
             nodes near coords,
             point meta=explicit symbolic]
    coordinates {
        (146.496, 0.295) [AXON]
    };
    
    \addplot[scatter,only marks,blue,
             nodes near coords,
             point meta=explicit symbolic]
    coordinates {
        (154.896, 0.406) [Brent-Kung]
        (161.952, 0.329) [Han-Carlson]
        (239.568, 0.273) [Kogge-Stone]
        (195.048, 0.429) [Sklansky]
    };
    
    \end{axis}
    \end{tikzpicture}
    \caption{Comparison of area and delay for (1) 32-bit adder generated by~\cite{Roy2014,Roy2016} and synthesized by \textit{Design Compiler} under different timing constraints, (2) 32-bit adder netlist generated by \textit{AXON}, and (3) four classical parallel-prefix adders.}
    \label{fig:compare}
    \vspace{-10pt} 
    \end{figure}
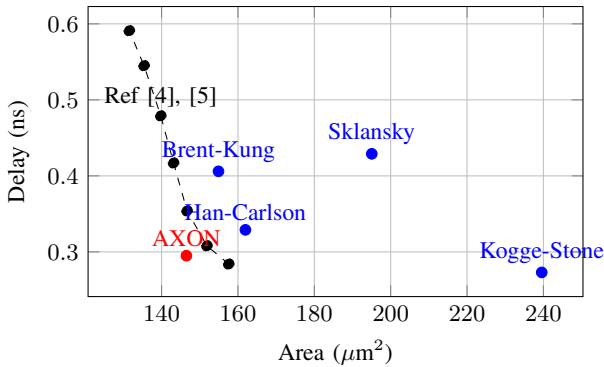


{We run full backend APR on the 5--20 selected designs and compare against two baselines: \textbf{Baseline 1} synthesizes a plain RTL ``A+B'' using commercial tools under a timing constraint; \textbf{Baseline 2} uses the minimal-node prefix topology from Step~1 (equivalent to~\cite{Roy2014,Roy2016}) in RTL without structural optimizations, again leaving synthesis/APR to the tools. Our method uses the full \textit{AXON} flow to generate and validate optimized netlists. Experiments use \textit{Synopsys Design Compiler} and \textit{Cadence Innovus} targeting \textit{TSMC 28nm}. We observe routing increases delay substantially (typically $>30\%$, and $>40\%$ for 32-bit), motivating physical-level evaluation.}

The final results are shown in Fig.~\ref{fig:ppa_delay_area}. For \textit{AXON}, we report only the design with the minimum ADP. \textit{AXON} demonstrates significant advantages over Baseline 1 and Baseline 2 across all metrics. It reduces delay by 9.9–24.2\% versus Baseline 1 and 1.4–10.3\% versus Baseline 2, while achieving comparable or slightly lower area, especially for larger designs. These gains lead to advantages in composite metrics: ADP improves by up to 12.6\%, and EDP decreases by up to 32.1\%, reflecting significantly better energy efficiency. These results confirm the superior efficiency of \textit{AXON} in balancing delay, power, and area, making it suitable for both high-performance and energy-constrained applications.

Moreover, the \textit{Design Compiler (DC)} results are highly sensitive to timing constraints. Although DC is powerful and can aggressively transform the RTL to meet timing (except under extremely tight constraints), the resulting designs do not necessarily achieve the best area–delay product. We synthesized the 32-bit minimal-node adders from \cite{Roy2014, Roy2016} under varying timing constraints, obtaining a set of designs that approximate a Pareto frontier (Fig.~\ref{fig:compare}). We then evaluated the netlist generated by \textit{AXON} together with four classical parallel-prefix adders. The result shows that \textit{AXON} surpasses this frontier and outperforms all four baselines, highlighting its PPA advantages.

\section{Conclusion}\label{sec:conclusion}

This work presents \textit{AXON}, an automated hierarchical adder design and optimization framework that explores the design space from architectural to netlist levels. By integrating the Prefix-Ling hybrid architecture, selective $P$-network and inverter insertion, and a two-stage evaluation methodology, \textit{AXON} achieves significant improvements in delay, area, and power while preserving functional equivalence. It is noted that \textit{AXON} aims at optimizing logic synthesis, and therefore it can be combined with existing works on prefix-topology exploration. Experimental results demonstrate the its efficiency against commercial EDA tools, providing a practical solution and paradigm for automated adder design and synthsis.

\bibliographystyle{IEEEtran}
\bibliography{references}

\end{document}